\pdfoutput=1

\documentclass[aps,prd,twocolumn,superscriptaddress]{revtex4}

\usepackage{amsmath}
\usepackage{amsfonts,amssymb}
\usepackage{graphicx,graphics}
\usepackage[hidelinks]{hyperref}
\usepackage{bm}

\usepackage{color}

\usepackage[normalem]{ulem}

\DeclareSymbolFont{symbolsC}{U}{pxsyc}{m}{n}
\DeclareMathSymbol{\coloneqq}{\mathrel}{symbolsC}{"42}

\begin{document}
	
\title{Homoclinic chaos in the Hamiltonian dynamics of extended test bodies}

\author{Ronaldo S. S. Vieira}\email[]{ronaldo.vieira@ufabc.edu.br}
\affiliation{Centro de Ci\^encias Naturais e Humanas, Universidade Federal do ABC, 09210-580 Santo Andr\'e, SP, Brazil}

\author{Ricardo A. Mosna}\email[]{mosna@unicamp.br}
\affiliation{Departamento de Matem\'atica Aplicada, Universidade Estadual de Campinas, 
	13083-859,  Campinas,  S\~ao Paulo,  Brazil}   

%
\begin{abstract}
There is a long tradition of studying chaotic trajectories in systems whose integrability is broken by means of an external perturbation. Here we explore a different route to chaos, in the dynamics of extended bodies, which arises due to finite-size corrections to the otherwise integrable motion of a test particle. We find that cyclic changes in the overall shape of the body may lead to the onset of chaos. This is applied to the Duffing and Yukawa potentials. For Kepler's potential, periodic deviations from spherical symmetry give rise to chaotic regions around the unperturbed parabolic orbit.
\end{abstract}


	
	
	

\maketitle
%
%
%
%
%

\section{Introduction}

Systems for which the objects under study can be modeled as small test bodies are the cornerstone of the physical sciences, from astrophysics to microscopic physics. When it is possible to approximate those test bodies by point particles, powerful tools of Hamiltonian dynamics may often be applied. In many instances, however, one must take into account the internal structure of the body and this may bring about substantial changes in the late-time dynamics of the system. 

Here we consider a novel example of this phenomenon in the simple case of a spherically symmetric body in classical mechanics, with the resulting qualitative change in the dynamics being the presence/absence of chaos due to finite-size effects. We show that the quadrupole tensor of such a body has the effect of perturbing its underlying test-particle Hamiltonian, thus generating a (perturbed) test-body Hamiltonian flow for the center-of-mass motion. Moreover, we show that the ensuing finite-size effects may drastically change the qualitative dynamics of the system, breaking its integrability. 


Specifically, we consider a spherical test body with a periodic oscillating radius or charge distribution (\emph{i.e.}, a pulsating ball) and show that, in this case, the potential for the extended body gets a correction due to the coupling of an inertia parameter to the Laplacian of the potential of the underlying point-particle Hamiltonian. For the Duffing potential with negative quadratic constant, we find chaotic behavior near the zero-energy curve when the oscillation of the ball is taken into account. We also find homoclinic chaos for a Yukawa central force field. This is done by using Melnikov's technique.

Finally, we consider Kepler's problem and show that time-periodic deviations from spherical symmetry of an extended test body lead to homoclinic chaos around the unperturbed parabolic, zero-energy orbit.

\section{Dynamics of extended bodies}
\label{sec:formalism}

Consider the motion of a test body subjected to a potential field $V({\bm x})$ in classical mechanics. Let $\rho({\bm x, t})$ be the body's charge density distribution
which couples to $V({\bm x})$  (be it a mass density for a gravitational potential, an electric charge density for an electric potential, etc). The total force $\bm{F}$ and torque $\bm{\tau}$ on the body are then given by 
\begin{equation} \label{eq:force}
	{\bm F} = - \int \rho\, \nabla V\, d^3 x\,,
\end{equation}
and ${\pmb \tau} = - \int \rho\,({\bm x} - {\bm z})\,\times\, \nabla V\, d^3 x$, where ${\bm z}={\bm z}(t)$ is the body's center of charge, which is also the point with respect to which the torque is calculated.
Let us assume, for simplicity, that the charge distribution $\rho$ of the body is proportional to its mass density (this is automatically the case when $V$ is a gravitational potential). The body's center of charge then coincides with its center of mass. The resulting equations of motion are therefore given by $m\, {\ddot{\bm{z}}}(t) = {\bm F}$ and ${ \dot{\bm{S}}} = {\pmb \tau}$, where $m$ and $\bm{S}$ are the total mass and intrinsic angular momentum (spin)  of the body.

Let $\overline{\nabla V}$ be the average value of $\nabla V$ as felt by the body,
$$
	\overline{\nabla V} = \frac{1}{q}\int \rho\, \nabla V\, d^3 x,
$$
with $q=\int \rho\, d^3 x$ its total charge.
We note from Eq.~(\ref{eq:force}) that the center-of-mass motion is given, in terms of $\overline{\nabla V} $, by
\begin{equation} \label{eq:potencialaverage}
	\ddot{\bm{z}}(t) =  - \frac{q}{m}\overline{\nabla V}\,.
\end{equation}
If the body is spherically symmetric and if $\nabla^2 V = 0$ (\emph{e.g.}, a vacuum gravitational or electric field) then the mean-value property of harmonic functions tells us that the gradient of the potential calculated at the center of mass is equal to $\overline{\nabla V}$, so that 
$\ddot{\bm{z}}(t)=  - \frac{q}{m}\nabla V (\bm{z}(t))$, exactly as in the point-particle case. 

The case of a non-harmonic potential ($\nabla^2 V \ne 0$) is more interesting. In this case, $\overline{\nabla V}$ must be given by $\nabla V (\bm{z}(t))$ plus an extra contribution that somehow should  be a function of $\nabla^2 V$. In fact, this extra contribution is exactly proportional to $\nabla^2 V$ in the case of spherical test bodies in the quadrupole approximation, for which the effective test-body potential is given by Eq.~(\ref{eq:potentialBall}). As a result, an effective Hamiltonian formulation for the center-of-mass motion of the body, which takes into account finite-size effects perturbatively, can be obtained.

\subsection{Quadrupole approximation}
\label{sec:quadrupole}

Let us assume that the test body is small, in the sense that $\nabla V$ in Eq.~(\ref{eq:force}) is well approximated by a second-order expansion around the center of mass $\bm{z}$. We can then write
\begin{equation}
	F_i = -q\,\partial_i\,V - \frac{1}{2} Q^{jk}\,\partial_j \partial_k (\partial_i V)\,,
\end{equation}
where $q$ is the total charge of the test body. The quadrupole tensor $Q^{jk}$ of the charge distribution, with respect to the center of mass, is given by $Q^{jk}(t) = \int \rho\,\left(x^j - z^j(t)\right)\left(x^k - z^k(t)\right)\,d^3x$.
The equations of motion do not restrict the dynamics of the quadrupole tensor in any way; an object can actively control its internal structure at will \cite{harte2021AcAau}. 
Implicit in this approximation is the assumption that tidal deformations due to the external force gradient along the body are neglected when compared to body's internal forces. In other words, we assume a strong enough internal mechanism in such a way that external tidal forces to not affect the body's structure.
We may then consider $Q^{jk} = Q^{jk}(t)$ as a prescribed function of $t$. As a result, the total force on the body may be written as 
\begin{equation}\label{eq:totalforce}
F_i = -q\,\partial_i\,\mathcal{V}\,,
\end{equation}
where
\begin{equation}\label{eq:potentialBody}
	\mathcal{V}(\bm{x},\,t) \coloneqq V(\bm{x}) + \frac{1}{2q} Q^{jk}(t)\,\partial_j \partial_k\,V(\bm{x})
\end{equation}
with both sides of Eq.~(\ref{eq:totalforce}) evaluated at the center of mass $\bm{z}(t)$.

As a result,
\begin{equation}
\ddot{\bm{z}}(t) = - \frac{q}{m}\nabla\,\mathcal{V}(\bm{z},\,t)\,.
\end{equation}
This shows that the dynamics of the center of mass in the quadrupole approximation is described by the effective test-body Hamiltonian (here taken with dimensions of energy per unit mass)
\begin{equation}\label{eq:testbodyHamiltonian}
	\mathcal{H} = \frac{1}{2}\,{\bm p}^2 + \frac{q}{m}\mathcal{V}\,,
\end{equation}
where $\bm{p} = d\bm{z}/dt$ is the specific linear momentum of the center of mass.

The torque on the body with respect to $\bm{z}(t)$ is given by 
$\tau_i = -\,\varepsilon_{ijk}Q^{jl}\partial_l \partial^k\,V$ \cite{harte2021AcAau}.
In order to analyze only finite-size effects related to the center-of-mass motion, we restrict ourselves to torque-free configurations so that we can consistently assume spin-free motion (more on this below).

\subsection{Small (spherically symmetric) balls}

If the body is spherically symmetric, then the expression for $Q^{ij}$ simplifies to $Q^{ij} = \mu\,\delta^{ij}$, where $\delta_{ij}$ is the Euclidean metric and
$\mu = \frac{1}{3}\int \rho\, |{\bm x} - {\bm z}|^2 \,d^3x$
is always positive. 
In this case, since charge and mass distributions are proportional, we have $\mu = I q/(6m)$, where $I$ is the trace of the body's inertia tensor. We will call $\mu$ the \emph{inertia parameter} of the body. 
We see that the effective $\mu$ is of order $\sim q\,R^2$, where $R$ is the characteristic radius of the body.

The effective test-body potential $\mathcal{V}$ then reduces to
\begin{equation}\label{eq:potentialBall}
	\mathcal{V}({\bm x},t) = V({\bm x}) + \frac{1}{2 q}\,\mu(t)\,\nabla^2\,V({\bm x})\,,
\end{equation}
and the total torque vanishes. 
For $\nabla^2\,V = 0$, the dynamics reduces to that of a test particle, even if the mass distribution changes with time (while maintaining spherical symmetry).

\section{Homoclinic chaos due to time-dependent quadrupole parameters}
\label{sec:Melnikov}

We summarize below the main results of Melnikov's method \cite{holmes1990PhysRep, lichtenbergLieberman1992} to detect chaos around an unperturbed homoclinic orbit when a small time-periodic perturbation is added to the system.
We restrict ourselves to one-degree-of-freedom systems described by an unperturbed Hamiltonian of the form
\begin{equation}
\mathcal{H}_0 = \frac{1}{2}p^2 + \frac{q}{m} V(x),
\end{equation}
where ($x$, $p$) are canonical coordinate and momentum variables. Note that this restriction does not require that the physical system is one dimensional, but only that it can be reduced to one effective degree of freedom after considering its symmetries.
For a thorough exposition of Melnikov's method and its extension to the non-Hamiltonian case, see Ref.~\cite{guckenheimerHolmes2013}.

\begin{figure*}
	\begin{center}
		\includegraphics[width=0.38\textwidth]{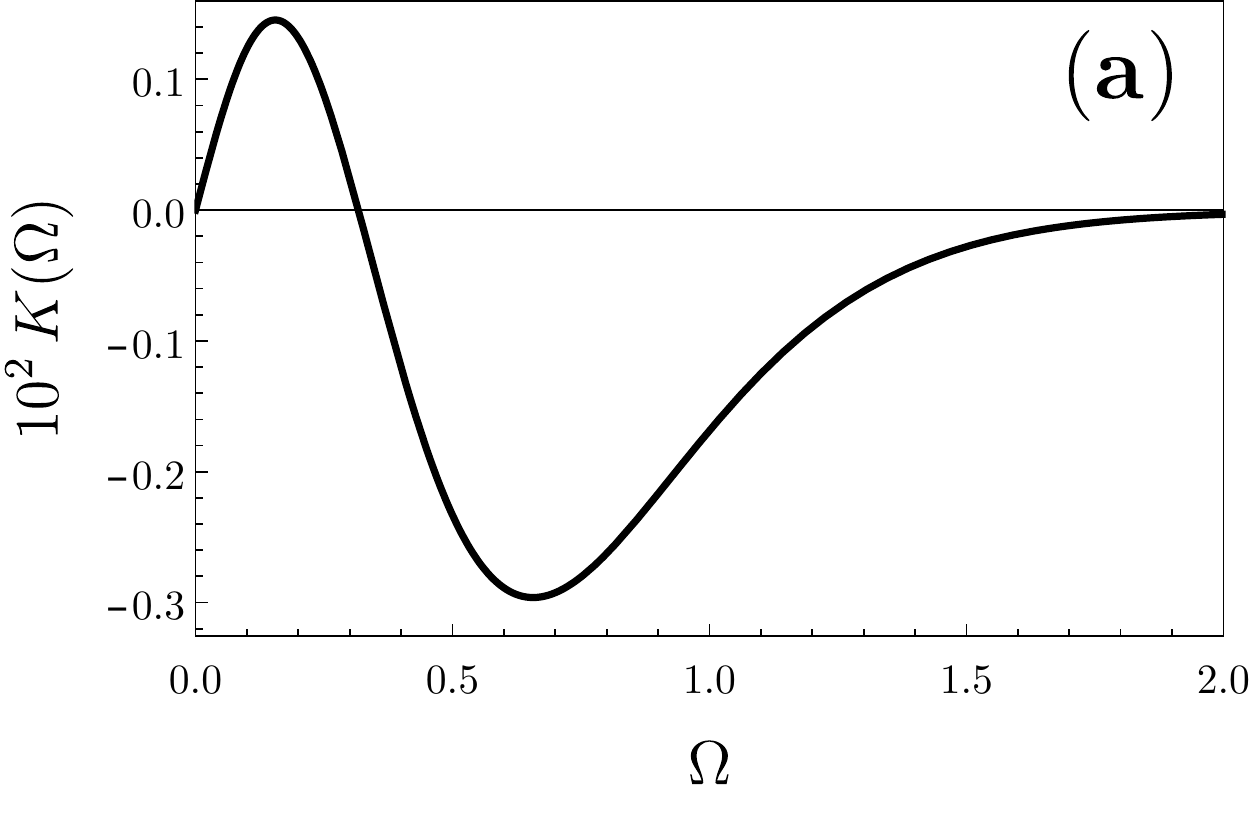} 
		\includegraphics[width=0.38\textwidth]{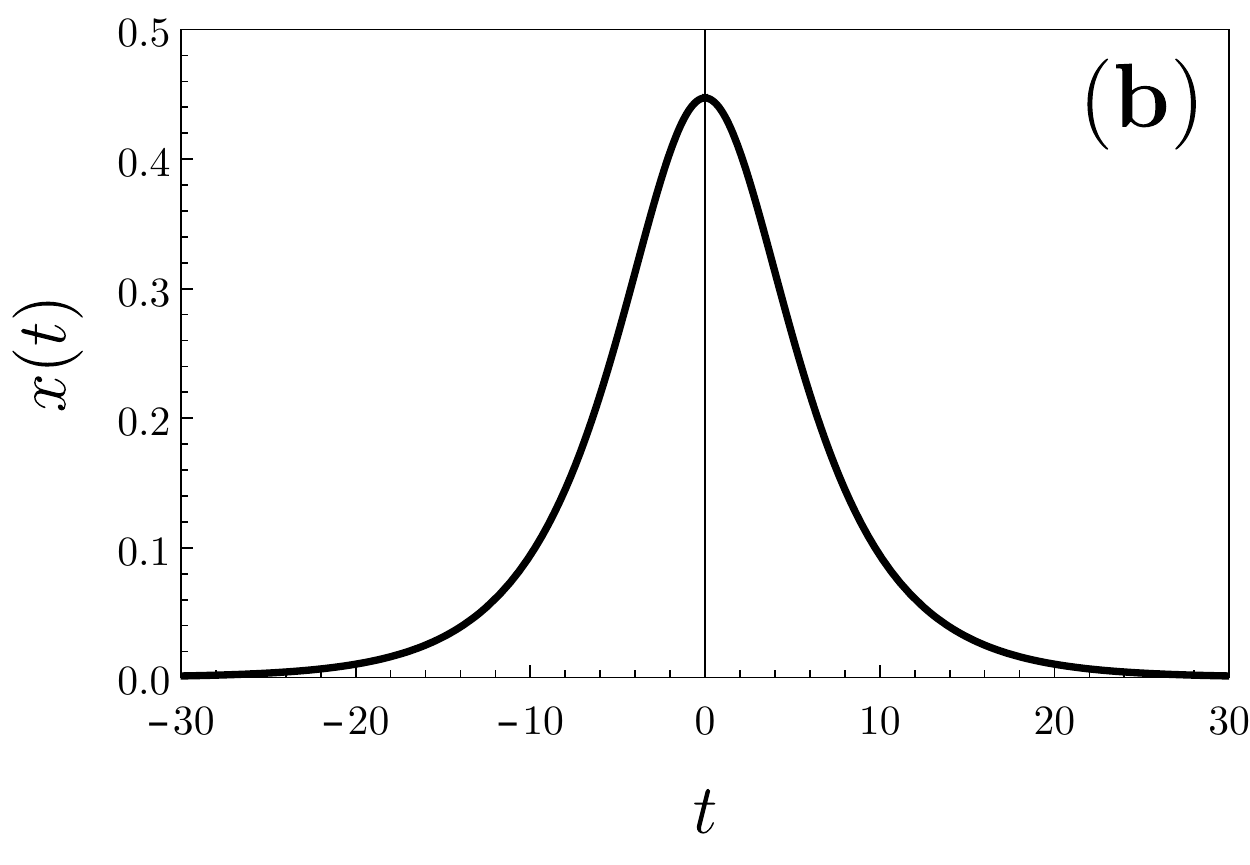}  	
		\includegraphics[width=0.38\textwidth]{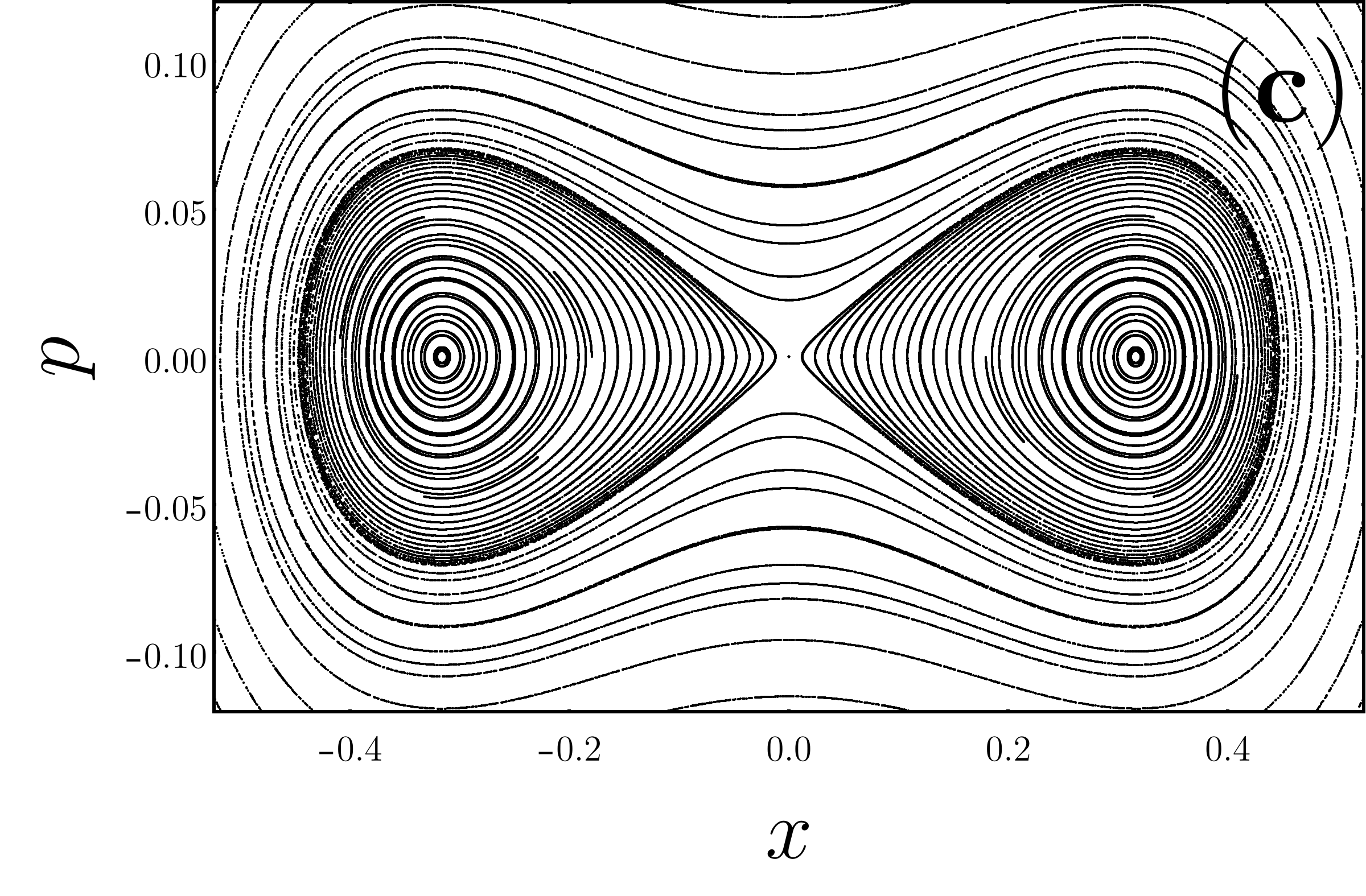} 
		\includegraphics[width=0.38\textwidth]{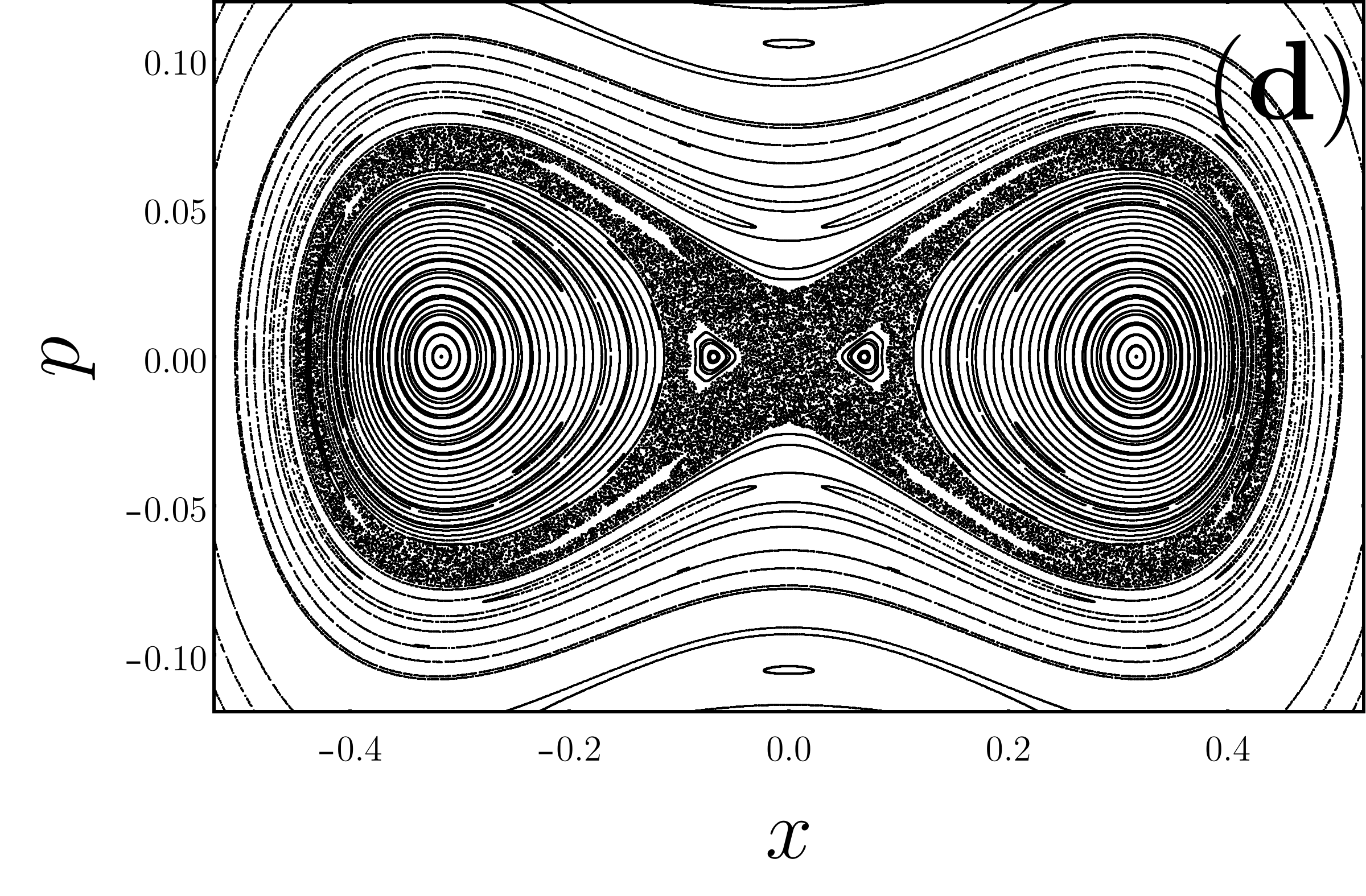}	
		\caption{
			(a) $K(\Omega)$ for a small ball subject to the Duffing potential, with $k = 0.1$ and $\omega=1$.
			(b) Corresponding plot of $x = x(t)$ for the unperturbed homoclinic orbit, with $t=0$ at the turning point. 
			(c) Phase-space portrait for point-particle motion in the Duffing potential, with $\mu = 0$ and parameters $k = 0.1$,  $w = 1$.
			(d) $x-p$ stroboscopic Poincar\'e section for the pulsating ball with $\mu_0 = 0.0005$ and $\Omega = 0.6$. The consequents are evaluated for integer multiples of the period of internal oscillation of the ball, $t_n = 2n\pi /\Omega$. The central region, which  encompasses the unstable fixed point, becomes chaotic.}
		\label{fig:KWduffing}
	\end{center}
\end{figure*}

We add to this Hamiltonian a time-dependent periodic perturbation $\epsilon\,\mathcal{H}_1(x, p, t)$ with frequency $\Omega$ and $\epsilon$ small. The corresponding ``stroboscopic'' Poincar\'e section in the extended phase space, characterized by the consequents on the $x-p$ surface defined by $t_n = 2n\pi/\Omega$, will present stable and unstable manifolds of the perturbed unstable fixed point. 
This perturbation may generate transversal intersections of these manifolds and thus give rise to homoclinic chaos \cite{lichtenbergLieberman1992, tabor1989chaos}.
One way to evaluate such transversal intersections is by means of the Melnikov integral \cite{lichtenbergLieberman1992, holmes1990PhysRep}
\begin{equation}
\mathcal{M}(t_0) = \int_{-\infty}^{\infty}\{\mathcal{H}_0\, , \mathcal{H}_1\}\, dt\,,
\end{equation}
with $\{\cdot\, , \cdot\,\}$ the Poisson brackets, $\mathcal{H}_0 = \mathcal{H}_0(x(t), p(t))$ and $\mathcal{H}_1 = \mathcal{H}_1(x(t), p(t), t+t_0)$, where ($x(t)$, $p(t)$) is evaluated along the unperturbed homoclinic orbit. The quantity $\mathcal{M}(t_0)$ is proportional to the first-order term in $\epsilon$ in the transversal distance (in phase space) between the unstable and stable manifolds, as measured with respect to the unperturbed homoclinic orbit, at $t=t_0$ \cite{holmes1990PhysRep}. Notice that $\mathcal{M}(t_0)$ is a periodic function of $t_0$. 
Simple zeros of the Melnikov integral correspond to transversal intersections of the stable and unstable manifolds and, therefore, imply homoclinic chaos (around the perturbed unstable fixed point of the stroboscopic Poincar\'e section).

Let us consider the finite-size structure of a small ball with time-dependent inertia parameter $\mu(t)$ being due to, for instance, an internal motion or mass redistribution of the test body. The body oscillates between a small ball with inertia parameter $2 \mu_0>0$ and a point particle,
\begin{equation}\label{eq:qcost}
\mu(t) = \mu_0 \big[1 + \cos (\Omega\, t) \big].
\end{equation}
If $\mathcal{H}_0$ is the point-particle Hamiltonian, then
\begin{equation}\label{eq:H1qt}
\epsilon \, \mathcal{H}_1 = \frac{1}{2q}\mu(t)\,\nabla^2 V(x),
\end{equation}
and therefore the Melnikov integral takes the form \cite{vieiraLetelier1998PhLA} 
\begin{equation}
\label{eq:epsM}
\epsilon \, \mathcal{M}(t_0) = -\frac{\mu_0}{2 q}\int_{-\infty}^{\infty} p(t)\, \frac{d}{dx} (\nabla^2 V)\,(1+\cos\big[\Omega(t+t_0)\big])\, dt
\end{equation}
with the integrand evaluated along the unperturbed homoclinic orbit $(x(t), p(t))$. This orbit is such that $x_{\text{uns}}\coloneqq x(\pm\infty)$ is the unstable equilibrium point and, if we choose the origin of $t$ such that $x_{\text{ret}}\coloneqq x(0)$ is the turning point, then $x(t)$ is an even function of $t$. As a result, by parity considerations, the Melnikov integral simplifies to 
\begin{equation}\label{eq:MKsin}
\epsilon \, \mathcal{M}(t_0) = \frac{\mu_0}{q} K(\Omega)\,\sin(\Omega\, t_0)\, ,
\end{equation}
with, considering the case of $p=dx/dt$,
\begin{equation}\label{eq:KOmega}
K(\Omega) = \int_{x_{\rm uns}}^{x_{\rm ret}}\frac{d\, (\nabla^{2} V)}{d x}\,\sin\big[\Omega\,t(x)\big]\, dx\, .
\end{equation}
Here, $-\infty <t(x)<0$ is evaluated along the unperturbed homoclinic orbit and $x$ denotes the coordinate of the already reduced 1D system.

Therefore, if $K(\Omega)\neq 0$ then  $\mathcal{M}$ possesses an infinite number of simple zeros, \emph{i.e.}, an infinite number of  transversal crossings of the unstable and stable manifolds associated with the unstable fixed point of the stroboscopic Poincar\'e section $t_n = 2n\pi/\Omega$. This implies homoclinic chaos \cite{lichtenbergLieberman1992}.

We now present two models for which finite-size effects generate chaos via the mechanism just discussed.
We take, without loss of generality,  $q=m=1$ (in appropriate units).

\subsection{Duffing potential with negative quadratic term}

The one-dimensional Duffing potential with negative quadratic term is given by
\begin{equation}\label{eq:quartic1D}
V(x) = -\frac{1}{2}\,k\,x^2 + \frac{1}{4}\,\omega\,x^4,
\end{equation}
with $k,\,\omega>0$.
It is well known for its applications in a wide rage of areas, such as electronics, optomechanics, micro and nanoresonators, etc~\cite{jin2014AppPL}. 

Consider the homoclinic orbit which corresponds to the positive-$x$ part of the $E=0$ energy level of the test-particle Hamiltonian, with the unstable fixed point at $x=0$ and the turning point  at $x=\sqrt{2 k/\omega}$. 
In this case, we have 
$d(\nabla^2 V)/dx = 6\omega x$.

Figure \ref{fig:KWduffing}(a) depicts the function $K(\Omega)$ for $k = 0.1$ and $\omega=1$. We have that $K(\Omega)$ is nonzero almost everywhere, so that $K(\Omega)=0$ for only a discrete set of values of $\Omega$. Therefore, this system presents homoclinic chaos around the perturbed fixed point of the $x-p$ stroboscopic Poincar\'e section for almost all values of $\Omega$, regardless of how small the perturbation is. Figure~\ref{fig:KWduffing}(b) shows the corresponding homoclinic unperturbed trajectory $x(t)$, which is an even function of $t$. We show in Fig.~\ref{fig:KWduffing}(c) the phase portrait of the test-particle motion and in 
Fig.~\ref{fig:KWduffing}(d) the $x-p$ stroboscopic Poincar\'e section for the motion of the corresponding small pulsating ball. We see that the pulsations clearly generate chaos around the central fixed point. The value of $\Omega$ was chosen such that $K(\Omega)\neq 0$.

\subsection{Yukawa potential}
\label{sec:2D}
Let us now consider the center-of-mass motion of a small spherical ball in a central potential $V(r)$.
The effective-potential formulation for the small-ball Hamiltonian (from Eqs.~(\ref{eq:testbodyHamiltonian}) and (\ref{eq:potentialBall}) with $q=m=1$) then gives rise to a (non-conserved) energy function
\begin{equation}\label{eq:energyCentral}
E=E(r,\dot r, t)= \frac{1}{2}\left(\frac{dr}{dt}\right)^2 + \mathcal{V}_{\rm eff}(r,t)\,,
\end{equation}
where
\begin{equation}\label{eq:VVeffball}
\mathcal{V}_{\rm eff}(r,t) = V(r) + \frac{L^2}{2 r^2} + \frac{1}{2}\,\mu(t)\,\nabla^2\,V
\end{equation}
and $L = r^2 d\varphi/dt$ is the body's (conserved) specific angular momentum.

Consider the Yukawa potential 
\begin{equation}\label{eq:PhiY}
V(r) = -\frac{k \, e^{-\lambda r}}{r}\,,
\end{equation}
with $k>0$ and $\lambda>0$. The effective potential (\ref{eq:VVeffball}) is then given by 
\begin{equation}
\mathcal{V}_{\rm eff}(r,t) = -\frac{k \, e^{-\lambda r}}{r} + \frac{L^2}{2 r^2} - \frac{1}{2}\,\mu(t)\,\lambda^2 \frac{k \, e^{-\lambda r}}{r}\,.
\end{equation}
Let us consider the case of $L\neq 0$ and perform the reparametrizations $t' = k^2 t/L^3$, $r' = k r/L^2$, $E' = L^2 E/k^2$, $\lambda' = \lambda L^2/k$ and $\mu' = k^5 \mu/L^8$. Then, after dropping the primes, Eq.~(\ref{eq:energyCentral}) remains valid with 
\begin{equation}\label{eq:VVeffY}
\mathcal{V}_{\rm eff}(r,t) = -\frac{e^{-\lambda r}}{r} + \frac{1}{2 r^2} - \frac{1}{2}\,\mu(t)\,\lambda^2 \frac{ e^{-\lambda r}}{r}\,.
\end{equation}
The form of the unperturbed effective potential ($\mu=0$) for different values of $\lambda$ is shown in Fig.~\ref{fig:V_yukawa}. It is not hard to show that this has a local maximum as long as $\lambda<\Lambda$, with  $\Lambda \cong 0.84$
{\footnote{The exact value is $\Lambda = \alpha e^{\beta} \left[1+W_{\!-1}\!\left(\alpha e^{\alpha}\right)\right]$, where $W$ is the Lambert $W$-function, $\alpha=-\frac{3+\sqrt{5}}{2}$ and $\beta=-\frac{1+\sqrt{5}}{2}$.}. This implies the existence of a homoclinic orbit in the ($r$, $p_r$) phase space whenever $\lambda\in(0,\Lambda)$. 
	\begin{figure}[ht]
	\centering
	\includegraphics[width=0.9\columnwidth]{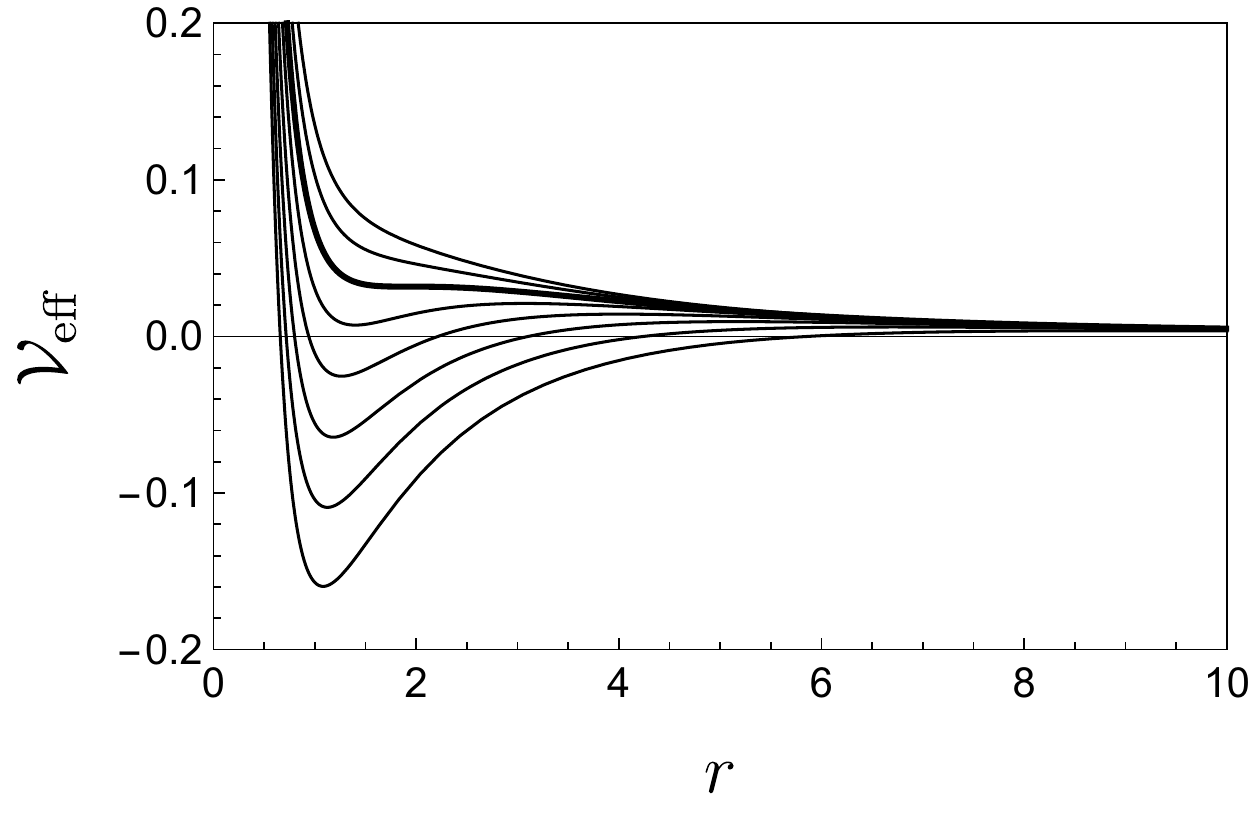}
	\caption{Yukawa effective potential for test-particle motion [Eq.~(\ref{eq:VVeffY}) with $\mu=0$] for evenly spaced values of $\lambda$ going from $\lambda= 0.5 \, \Lambda$ (lowest curve) to $\lambda= 1.2 \, \Lambda$ (uppermost curve), with $\Lambda \cong 0.84$. The thick line corresponds to $\lambda= \Lambda$, which is the largest positive $\lambda$ for which the effective potential has a local maximum (and hence an unstable circular orbit).
	}
	\label{fig:V_yukawa}
	\end{figure}


For a spherically symmetric (small-ball) perturbation of the form (\ref{eq:qcost}), Eqs. (\ref{eq:MKsin}) and (\ref{eq:KOmega}) yield
\begin{equation}
K(\Omega) = -\frac{\lambda^2}{2}\int_{r_{\rm ret}}^{r_{\rm uns}}
\frac{e^{-\lambda r}}{r}\bigg(\lambda+\frac{1}{r}\bigg)
\,\sin\big[\Omega\,t(r)\big]\, dr,
\end{equation}
with $0<t(r)<\infty$ evaluated along the unperturbed homoclinic orbit, for which the energy equals the value of the effective potential evaluated at its local maximum $r_{\rm uns}$. We adopt $t=0$ at the turning point (notice that $r_{\rm ret}<r_{\rm uns}$ in this case). 

Going back to the original variables, we see that the unperturbed potential admits an unstable fixed point as long as $0<\lambda L^2/k<\Lambda\cong 0.84$. Therefore, there is a corresponding homoclinic orbit in this system for any value of the (original) Yukawa parameter $\lambda>0$, provided we restrict  the body's angular momentum to the range $L^2< k\,\Lambda/\lambda$.

We illustrate in Fig.~\ref{fig:KW}(a) how $K(\Omega)$ typically depends on $\Omega$ and in Fig.~\ref{fig:KW}(b) the corresponding homoclinic unperturbed trajectory $r(t)$. We see that $K(\Omega)$ is different from zero almost everywhere in the range of frequencies shown in the plot. Therefore, from the discussion of the previous paragraph, a pulsating test body does give rise to chaotic dynamics in this case, regardless of its size.

\begin{figure*}[t]
	\begin{center}
		\includegraphics[width=0.38\textwidth]{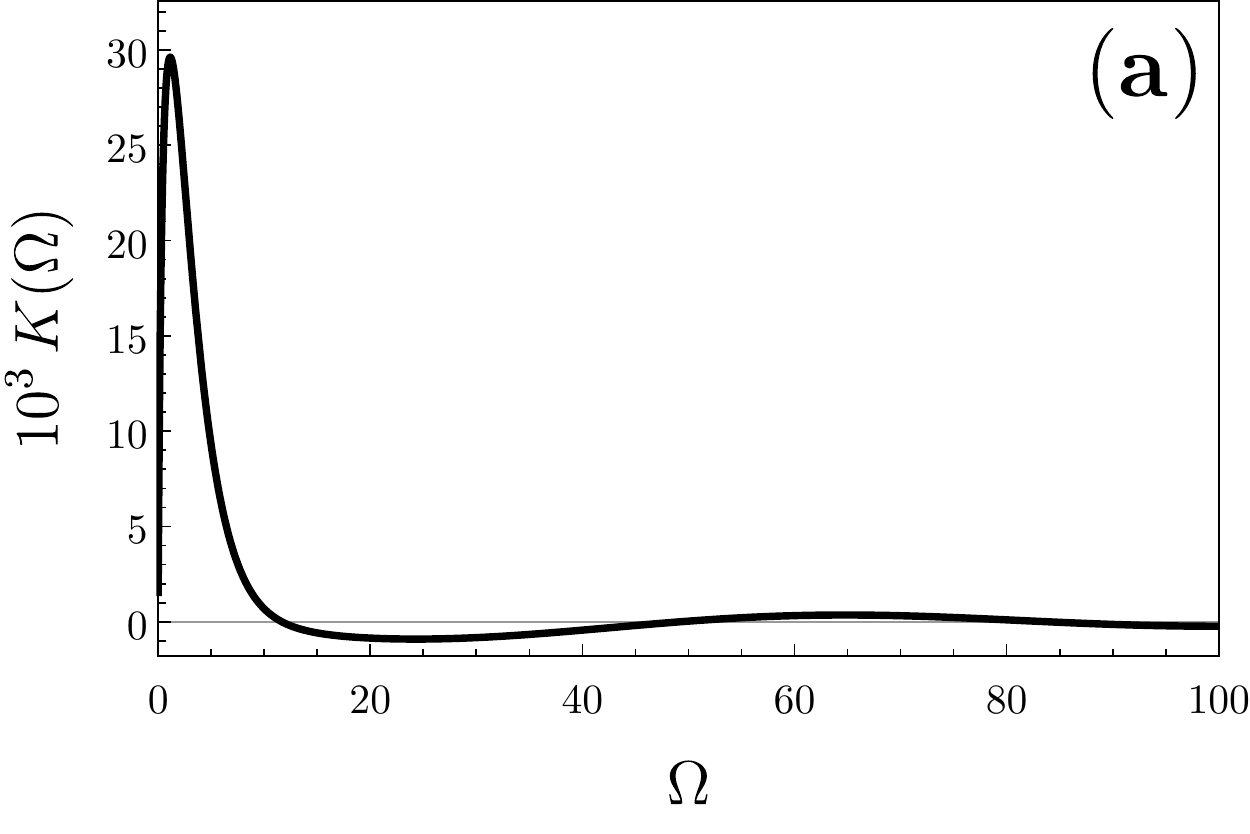}
		\includegraphics[width=0.38\textwidth]{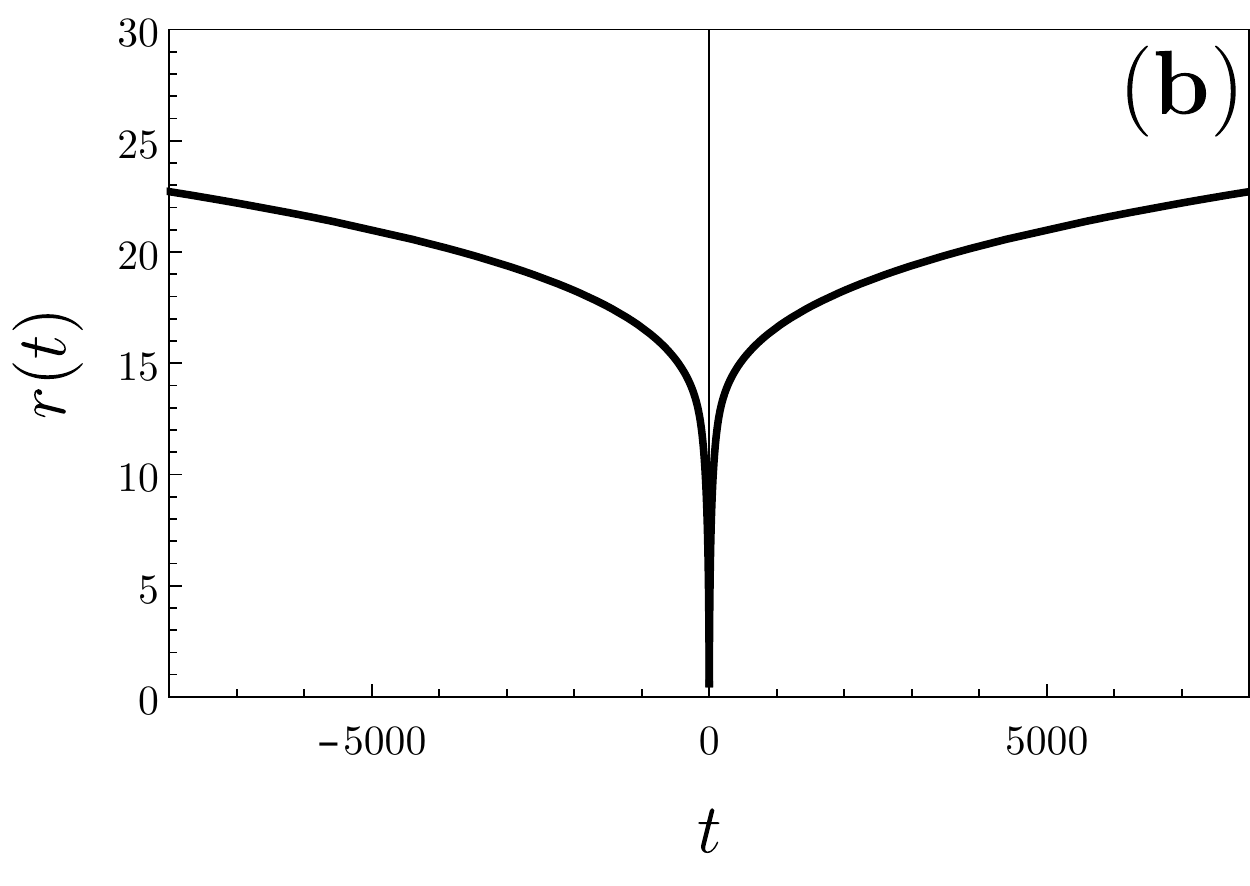}	
		\includegraphics[width=0.38\textwidth]{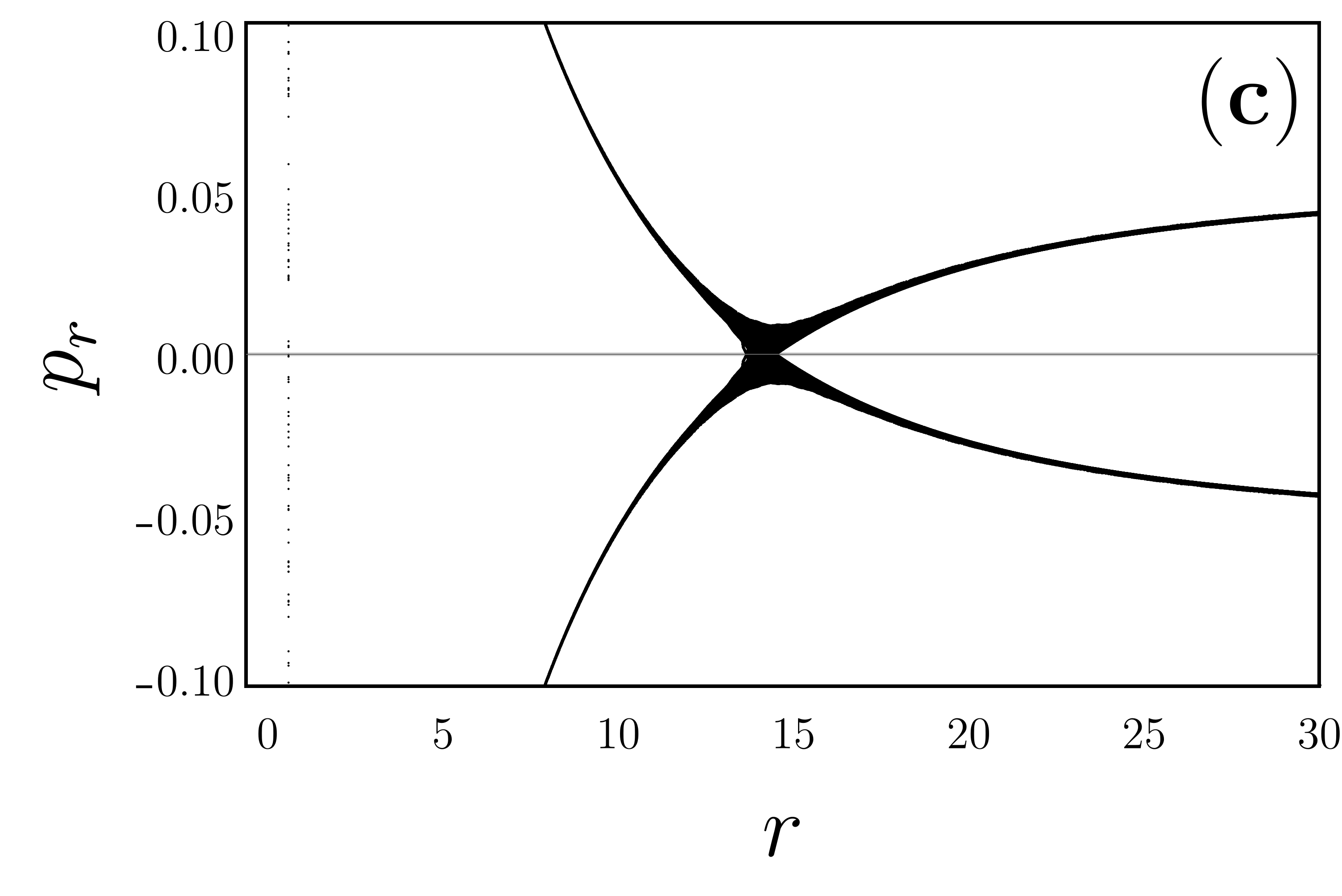}
		\includegraphics[width=0.38\textwidth]{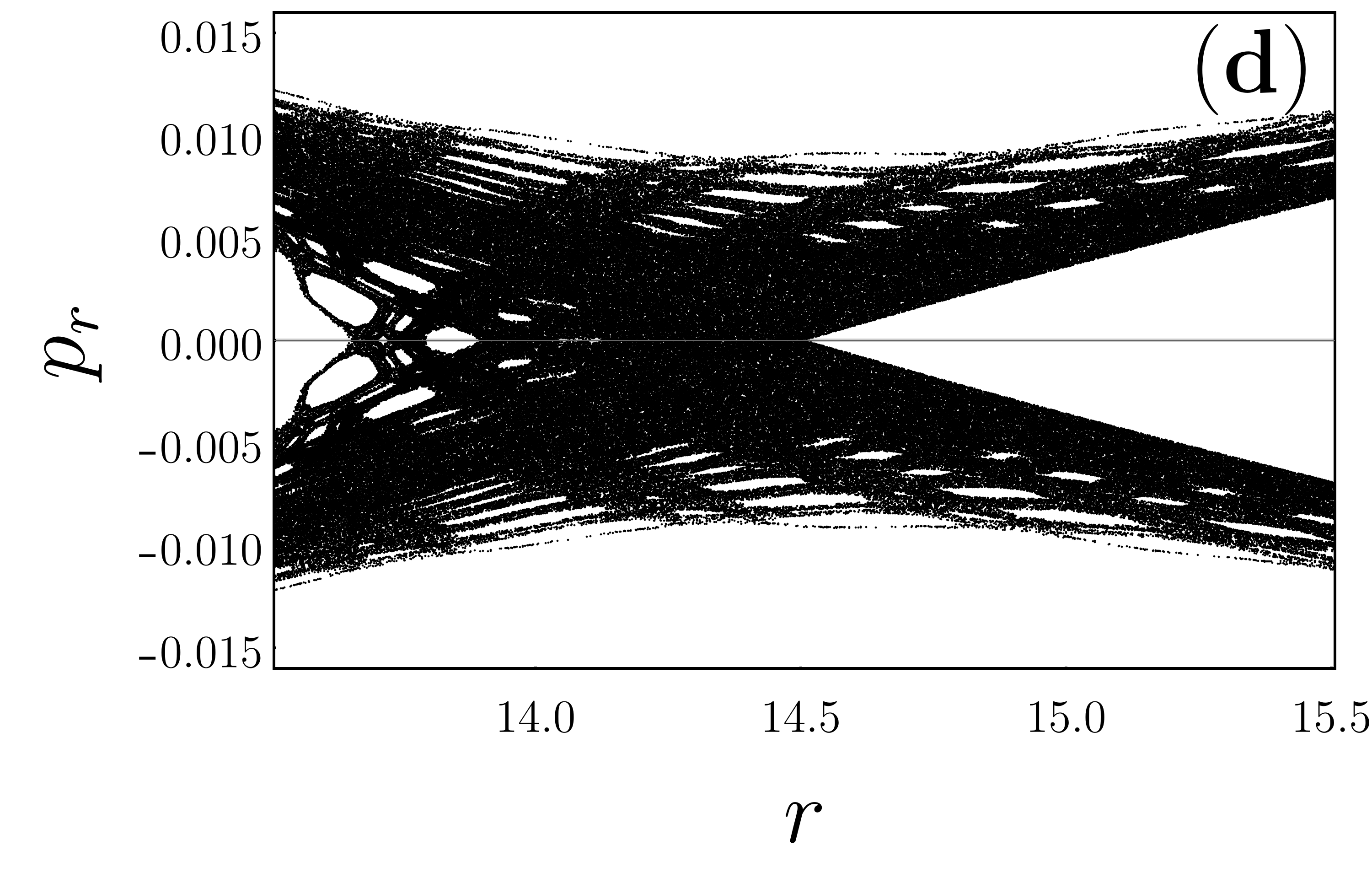}
		\caption{
		(a) $K(\Omega)$ for a small ball subject to the Yukawa potential, with $\lambda=0.3$. 
		(b) Corresponding plot of $r = r(t)$ for the unperturbed homoclinic orbit, with $t=0$ at the turning point. 
		(c) Consequents for one orbit in the stroboscopic Poincar\'e section for the Yukawa potential with an infinite barrier added at $r_{\rm bar} = 30$. The initial conditions are $r_0 = r_{\rm un} - 0.10$ and $p_{r, 0}=0$, with the system parameters given by $\lambda = 0.3$, $\mu_0 = 0.0001$ and $\Omega = 0.1$. The unperturbed unstable fixed point is given by $r_{\rm un} = 14.5071$, $p_{r,\rm{un}} = 0$. 
		(d) Zoom near the unperturbed unstable fixed point. 
		}
		\label{fig:KW}
	\end{center}
\end{figure*}

In practice, we may confine the Yukawa potential to a finite region in space (\emph{e.g.}, by including an outer potential barrier), so that the orbit does not escape to infinity but collides elastically with a confining wall.  In this way, the transverse homoclinic intersections are guaranteed by Melnikov's method in its strictly Hamiltonian version~\cite{holmes1990PhysRep}. Figs.~\ref{fig:KW}(c) and \ref{fig:KW}(d) show a stroboscopic Poincar\'e section for this system, which clearly exhibits chaos.


\section{Nonspherical shapes and the Kepler problem}

Kepler's potential has zero Laplacian and thus spherically symmetric bodies behave as point particles under its flow. One may, however, apply the formalism of the previous sections to a test body which is not spherically symmetric, so that the extra term in Eq.~(\ref{eq:potentialBody}) is nonzero even for a $V({\bm x})$ that is harmonic (\emph{i.e.}, $\nabla^2V=0$).

In order to keep the torque-free assumption, the body's center-of-mass position vector ${\bm{z}}$ must be an eigenvector of the quadrupole tensor $Q^{ij}$ \cite{harte2021AcAau}. 
Let $Q^{ij}_T$ be the traceless part of $Q^{ij}$, so that $(Q_T)^i_{\phantom{i} j}z^j = \alpha\, z^i$.
We choose the test body to be axially symmetric and such that its center of mass is located at the equatorial plane, so that $Q^{ij}_T$ is diagonal in $(R, \varphi,z)$ cylindrical coordinates. This yields $(Q_T)^R_{\ R} = (Q_T)^\varphi_{\ \varphi} = \alpha$ and $(Q_T)^z_{\ z} = -2\alpha$. A positive $\alpha$ corresponds to an oblate deformation, whereas a negative $\alpha$ corresponds to a prolate deformation. The test-body effective potential is then given by (\ref{eq:potentialBody}).

After the reparametrization which leads to Eq.~(\ref{eq:VVeffY}) (here for the Kepler case, $\lambda=0$), we choose a quadrupole time-periodic perturbation
\begin{equation}
\alpha(t) = \alpha_0\,\cos(\Omega\, t),
\end{equation}
with $\alpha_0>0$. This corresponds to an spheroid that is almost a sphere and which oscillates according to an alternating oblate-prolate periodic deformation. The perturbation to the effective potential then reads [see Eq.~(\ref{eq:potentialBody})]
\begin{equation}
	-\frac{3}{2 r^3}\alpha(t).
\end{equation}

If we take the unperturbed configuration to be the parabolic orbit with $E=0$, an extension of Melnikov's method may be applied to the homoclinic orbit associated with the unstable point at infinity. Interestingly enough, this dynamical system was shown to present homoclinic chaos via application of this extended Melnikov's method \cite{letelierMotter1999PRE}, although in a different physical context and with a different interpretation. 

Therefore, although a spherical test body (pulsating or not) always follows the well-known conics when subjected to the Kepler's potential field, the nonradial motion of an oblate-prolate changing-shape test body does exhibit chaos (around the parabolic unperturbed orbit), regardless of how small the perturbation is.


\section{Discussion}
\label{sec:Discussion}

The mean-value theorem implies that spherical (extended) bodies subjected to a harmonic potential (\emph{i.e.}, $\nabla^2 V=0$) have the same dynamics as point particles. However, as discussed in the main text, this is no longer true if any of these two requirements (spherical symmetry of the body or harmonicity of the potential) fail to be met. In this paper, we showed that this difference in the dynamics, brought about by finite-size effects of the test body, is enough to produce chaos in physical systems which are perfectly integrable in the point-particle limit.

To illustrate the first point we considered a small pulsating (spherical) ball as a test body in the quadrupole approximation. The pulsating ball is then subjected to the 1D Duffing potential and to the 3D Yukawa potential, which are not harmonic. We have then shown, via Melnikov's method, that the extended-body dynamics exhibits homoclinic chaos in both of these cases.  To illustrate the second point, we considered the case of Kepler's potential with a body having its shape oscillating between a prolate and an oblate spheroid. We showed that the mathematical problem is then the same as an example of homoclinic chaos already presented, in a different physical context, in Ref.~\cite{letelierMotter1999PRE}. Therefore, this system also presents homoclinic chaos.

One may generalize our results and assert that if a sufficiently small local maximum of $V$ is inside a potential well (so that motion around that point is bounded), a generic time-dependent internal redistribution of charge of a test body inside the well will break the integrability of its translational dynamics, leading to chaotic behavior around the fixed point. 
These effects may have applications in the semiclassical motion of Bose-Einstein condensates moving inside optical traps, once the finite size of the condensate is taken into account (see, for instance, Ref.~\cite{fujiwaraEtal2018NJP} and references therein).
Finally, it would be interesting to investigate if this effect could have astrophysical implications, for instance, in the motion of pulsating stars.

%

%

\section*{Acknowledgements} 
We acknowledge stimulating discussions with Fernanda F. Rodrigues.
R.A.M. was partially supported by Conselho Nacional de Desenvolvimento Cient\'{i}fico e Tecnol\'{o}gico under grant 310403/2019-7.

%




\end{document}